\documentstyle[11pt,epsf,fleqn]{elsart}

\newcommand{\be}[1]{\begin{equation}}
\newcommand{\ee}{\end{equation}}
\newcommand{\ba}[1]{\begin{eqnarray} \label{(#1)}}
\newcommand{\ea}{\end{eqnarray}}
\newcommand{\nn}{\nonumber}
\newcommand{\rf}[1]{(\ref{(#1)})}
\def \Rpv{R_{P} \hspace{-1.2em}/\;\hspace{0.2em}}
\def \Rpvv{R_{P} \hspace{-0.9em}/\;\hspace{0.2em}}

\def \znbb {0\nu\beta\beta}

\def \rp{$R_p \hspace{-0.9em}/\;\:\hspace{-0.1em}$}
\def \Rp{$R_p \hspace{-0.9em}/\;\:\hspace{-0.1em}$}

\begin{document}
\begin{frontmatter}
\title{{\bf Improved bounds on SUSY accompanied
neutrinoless double beta decay}}
\author{H. P\"as\thanksref{e1}}, 
\author{M. Hirsch\thanksref{e2}},
\author{H.V. Klapdor--Kleingrothaus\thanksref{e3}}
\address{
Max--Planck--Institut f\"ur Kernphysik,\\ 
P.O.Box 10 39 80, D--69029 Heidelberg, Germany}
\thanks[e1]{E-mail: Heinrich.Paes@mpi-hd.mpg.de}
\thanks[e2]{Present Address: {\it Inst. de Fisica Corpuscular - C.S.I.C.
- Dept. de Fisica Teorica, Univ. de Valencia, 46100 Burjassot, Valencia, 
Spain}, \\
E-mail: mahirsch@flamenco.ific.uv.es}
\thanks[e3]{E-mail: klapdor@mickey.mpi-hd.mpg.de}
\begin{abstract}
Neutrinoless double beta decay ($\znbb$)
induced by light Majorana neutrino
exchange between two decaying nucleons with squark/slepton exchange inside 
one and W exchange inside the other nucleon 
(so--called vector--scalar exchange)
gives stringent limits on 
R--parity violating interactions.
We have extended previous work by including the tensor contribution 
to the transition rate. We discuss the improved limits on trilinear
\Rp -MSSM couplings
imposed by the current experimental limit on the $\znbb$ decay half--life
of $^{76}$Ge.
 
\vspace{5mm}
\end{abstract}
\end{frontmatter}

In supersymmetric models, 
the new SUSY partners differ from the SM field content
in a discrete multiplicative quantum number R--parity defined as
\be{susy1}
R_{p}=(-1)^{3B+L+2S}.
\ee
Here $B$ denotes the baryon number, $L$ the lepton number and 
$S$ the spin of a 
particle leading to $R_p=+1$ for SM particles and $R_p=-1$ for superpartners.
Thus in $R_p$ conserving models superpartners can only be produced in pairs
and the LSP is stable, leading to a natural WIMP dark matter candidate.
While in the minimal supersymmetric extension (MSSM) of the standard model
(SM) R--parity
is assumed to be conserved, there are no theoretical reasons for $R_p$ 
conservation and
several GUT \cite{rpguts} and Superstring \cite{rpss} models require
R--parity violation in the low energy regime. 
Also the reports concerning an 
anomaly at HERA in the inelastic 
$e^+p$ scattering at high $Q^2$ and $x$ 
\cite{HERA} have renewed the interest in $\Rpv$--SUSY
(see for example \cite{kalino,drei}). 
Generally, one can add the following trilinear
R--parity violating terms to the
superpotential \cite{rpguts}
\be{susy2}
W_{\Rpvv}=\lambda_{ijk} L_i L_j \overline{E_k}+\lambda^{'}_{ijk}L_i Q_j 
\overline{D}_k +\lambda^{''}_{ijk}\overline{U}_i\overline{D}_j\overline{D}_k,
\ee  
where $i,j,k$ denote generation indices, $L,Q$ denote lepton and quark doublet 
superfields and $\overline{E},\overline{U},\overline{D}$ lepton, up-- and down 
quark singlet superfields. Terms proportional to $\lambda$, $\lambda^{'}
$
violate lepton number, those proportional $\lambda^{''}$ violate baryon 
number. 
While simultaneous presence of both kinds of terms would 
lead to too fast proton decay and thus is forbidden, 
assuming the $\lambda^{''}$ terms to be zero no constraints on 
$\lambda$ and $\lambda^{'}$ terms  can be derived from proton decay.

The search for 
neutrinoless double beta decay, converting a nucleus $(Z,A)$ into a nucleus 
$(Z+2,A)$ under emission of two electrons, has been proven to belong to
the most powerful tools to search for lepton number violating physics
beyond the SM (for a review see \cite{Kla97,Pascos}). 
Contributions occuring through Feynman graphs 
involving the exchange of superpartners as well as $\Rpv$--couplings 
$\lambda^{'}$ have been discussed in \cite{mohorg,Hir95,hir95d,Hir96,Faes}
and yield the most stringent bound on $\lambda^{'}_{111}$.
Taking into account the fact that the SUSY partners of the
left and right--handed quark states can mix with each other, also diagrams 
appear in which the neutrino--mediated double beta decay is accompanied by
SUSY exchange in the vertices (see fig. 1). 
These contributions allow to constrain also
combinations of couplings of higher generations 
$\lambda_{11j}^{'} \lambda_{1j1}^{'}$. 
They have been discussed in \cite{babu95} and more extensively 
in \cite{Hir96}, where however only scalar-pseudoscalar currents 
have been taken into account, whereas the tensor 
contribution to the decay rate has been neglected. On the other 
hand, in a recent work
\cite{Paes98a} the dominance of the tensor contribution in a
general framework for neutrino mediated double beta decay has been proven. 
In the present letter we reanalyze SUSY--accompanied double beta decay 
and discuss the relative importance of the different nuclear matrix 
elements involved.

The mixing between scalar superpartners $\tilde f_{L,R}$ of the left 
and right-handed fermions $f_{L,R}$ occurs due to non-diagonality 
of the mass matrix which can be written as 
\ba{e4}
{\cal M}^2_{\tilde f} &= \mbox{$ \left( \begin{array}{cc}
m^2_{\tilde{f}_L} + m_f^2 - 0.42 D_Z & - m_f (A_f + \mu \tan\! \beta) \\
- m_f (A_f + \mu \tan\! \beta) & m^2_{\tilde{f}_R} + m_f^2 - 0.08 D_Z
\end{array} \right) $}. 
\ea
Here,  $ f = d,  s,  b,  e,  \mu,  \tau$  and $\tilde f$ are 
their superpartners. $D_Z = M_Z^2 \cos\! 2\beta$ with 
$\tan\!\beta =\langle H_2^0\rangle / \langle H_{1}^{0} \rangle$ 
being the ratio of vacuum 
expectation values of the two Higgs doublets, $m_{\tilde f_{L,R}}$ 
are soft sfermion masses, $A_f$  are soft SUSY breaking parameters 
describing the strength of trilinear scalar interactions, and $\mu$ 
is the supersymmetric Higgs(ino) mass parameter. Once sfermion  
mixing is included, the current eigenstates $\tilde{f}_L, \tilde{f}_R$
become superpositions of the mass eigenstates $\tilde{f}_i$ with 
the masses $m_{\tilde f_i}$ and the corresponding  mixing angle 
$\theta^f$ is defined as

\ba{mixing}\nn
&&\sin\! 2\theta^f = \frac{2 m^2_{(f)LR}}
{m^2_{\tilde{f}_1}-m^2_{\tilde{f}_2}}, 
\ea
\ba{eigen} 
m^2_{\tilde f_{1,2}}=\frac{1}{2}
\Big(m^2_{LL} + m^2_{RR}\mp \sqrt{(m^2_{LL}-m^2_{RR})^2 + 4 m^4_{LR}}\Big) 
\ea
where $m^2_{LR}, m^2_{LL}, m^2_{RR}$ denote the $(1,2), (1,1), (2,2)$ 
entries of the mass matrix (\rf{e4}).

Now it is straightforward to find the  effective 4-fermion 
$\nu-u-d-e$ vertex induced by the sfermion exchange in the 
diagrams presented in fig. 1. The corresponding effective 
Lagrangian, after a Fiertz rearrangement, takes the form
\ba{L_SUSY}\nn
{\cal L}_{SUSY}^{eff}(x) &=&
\frac{G_F}{\sqrt{2}} \left[ \frac{1}{4} 
\left(\eta_{(q)LR}^{nj}  - 4 \eta_{(l)LR}^{nj} \right)\cdot U^{*}_{ni} 
\cdot \left(\bar \nu_i (1 + \gamma_5) e^c_j\right) 
\left(\bar u (1 + \gamma_5)  d\right)   - \right. \\ 
&-& 2 \eta_{(l)LL}^{nj} \cdot U_{ni}  
\cdot \left(\bar \nu_i (1 - \gamma_5) e^c_j\right) 
\left(\bar u (1 + \gamma_5)  d\right) + \\ \nn
&+& \frac{1}{2}\eta_{(q)RR}^{nj}\cdot U_{ni} 
\left(\bar \nu_i\ \gamma^{\mu} (1 + \gamma_5)  e^c_j\right) 
\left(\bar u\   \gamma_{\mu} (1 - \gamma_5)  d\right)  + \\ \nn
&+& \left. \frac{1}{8}\eta_{(q)LR}^{nj}\cdot U^{*}_{ni}  \cdot 
\left(\bar \nu_i\ \sigma^{\mu\nu}  (1 + \gamma_5)  e^c_j\right) 
\left(\bar u\   \sigma_{\mu\nu} (1 + \gamma_5)  d\right)\right].
\ea 
The \rp MSSM  parameters $\eta$ and neutrino mixing matrix 
$U_{ij}$ are defined as follows

\ba{eta}
\eta_{(q)LR}^{nj} &=& \sum_{k} \frac{\lambda'_{j1k}\lambda'_{nk1}}{2
\sqrt{2} G_F} 
\sin{2\theta^{d}_{(k)} }\left( \frac{1}{m^2_{\tilde d_1 (k)}} -  
\frac{1}{m^2_{\tilde d_2 (k)}}\right), \\
\eta_{(q)RR}^{nj} 
&=& \sum_{k} \frac{\lambda'_{j1k}\lambda'_{n1k}}{2 \sqrt{2} G_F}
\left( \frac{\sin^2{\theta^{d}_{(k)} }}{m^2_{\tilde d_1 (k)}} +  
\frac{\cos^2\theta^{d}_{(k)}}{m^2_{\tilde d_2 (k)}}\right), \\
\eta_{(l)LR}^{nj} &=&\sum_{k} \frac{\lambda'_{k11}\lambda_{njk}}
 {2 \sqrt{2} G_F}
 \sin{2\theta^{e}_{(k)} }\left( \frac{1}{m^2_{\tilde e_1 (k)}} -  
\frac{1}{m^2_{\tilde e_2 (k)}}\right), \\
\eta_{(l)LL}^{nj}&=&  \sum_{k} \frac{\lambda'_{k11}\lambda_{nkj}}
 {2 \sqrt{2} G_F}
\left(\frac{\cos^2{\theta^{e}_{(k)}}}{m^2_{\tilde e_1 (k)}} +  
\frac{\sin^2\theta^{e}_{(k)}}{m^2_{\tilde e_2 (k)}}\right),\\
\label{(neutr1)}
\nu^0_i &=& \sum_{j} U_{ij} \nu_j.
\ea
Here $\eta_{(f)LR}$ denotes the contribution vanishing in the 
absence of $\tilde f_L-\tilde f_R$ - mixing while $\eta_{(f)LL}$ 
and $\eta_{(f)RR}$ in this limit correspond to the $\tilde f_L$ and 
$\tilde f_R$ exchange contribution in fig. 1. We use the notations 
$d_{(k)} = d, s, b$ and $e_{(k)} = e, \mu, \tau$. Due to the antisymmetry 
of the Yukawa coupling $\lambda_{njk}$ in $nj$ it follows that 
$\eta_{(l)LR}^{nn} = 0$. This is an essential difference between 
the slepton $\tilde l_L-\tilde l_R$ and the squark 
$\tilde q_L-\tilde q_R$ contributions. The latter is not imposed 
to vanish at any combination of indexes.    
Since the $\eta_{(l)LL}^{nj}$  and  $\eta_{(q)RR}^{nj}$ contributions
to the diagram fig. 1 are helicity suppressed 
$\propto \langle m_{\nu} \rangle \simeq 
{\cal O}(0.5)$eV, we can restrict ourselves to the consideration
of 
$\eta_{(q)LR}^{nj}$  and  $\eta_{(l)LR}^{nj}$, which contributions
are momentum enhanced $\propto q \simeq p_F \simeq 100$ MeV,
where $p_F$ denotes the Fermi momentum.
Here $\langle m_{\nu} \rangle$ and $q$ denote the effective neutrino mass and 
momentum entering the
neutrino propagator. The first line in eq. \rf{L_SUSY} corresponds to the
scalar-pseudoscalar ($S+P$) contribution, the last line in \rf{L_SUSY} to 
the tensor contribution.

In the light neutrino case 
the $0\nu\beta\beta$ decay rate is given by 
\be{t12}
T^{-1}_{1/2}(\znbb) = G_{01}\left\{ 4 \bar\eta_{(l)} {\cal M}_{S+P} 
 +\left(- \bar\eta_{(q)}  + \eta_{(q)}\right)
\left({\cal M}_{S+P}+{\cal M}_{T}\right)
\right\}^2.
\label{t12}
\ee
Here $G_{01}$ denotes the phase space factor defined in \cite{Doi85},
$\eta_{(q)} = \eta_{(q)LR}^{11}$
and effective parameters are introduced as 
\ba{efpar}
\bar\eta_{(l,q)} = \sum_n\Delta_n \eta_{(l,q)LR}^{n1}.
\ea
For the $\bar\eta_{(l)}$ summation starts from $n=2$ and
$\Delta_n$ denotes the combination of mixing matrices corresponding
to heavy or sterile neutrinos
\be{delta}
\Delta_i = \sum''_{j} U^*_{ij} U_{ei} +   U^*_{is} U_{ei},
\ee
where the sum extends over heavy mass eigenstates $m_{\nu_i}> 10$ GeV
(see \cite{Hir96}).
In s-wave approximation for the outgoing electrons and under some assumptions 
according to \cite{tom91,Hir96} 
(the s-wave approximation is expected to affect the result less than 10 \%
\cite{Paes98a})
the matrix element is
\ba{me1}
{\cal M}_{S+P}
&=&\frac{F^{(3)}_{P}(0)}{4 R m_e G_A}\Big({\cal M}_{T^{'}}+\frac{1}{3}
{\cal M}_{GT^{'}}\Big),\\
{\cal M}_{T}&=&\alpha_1 \Big(\frac{2}{3} {\cal M}_{GT^{'}}-
{\cal M}_{T^{'}}\Big),
\ea
with (summation over nucleons $a,b$ is suppressed) 
\ba{matr01}
{\cal M}_{GT^{'}}&=&
\langle 0_f^+| h_{R}(\vec{\sigma_a}
\vec{\sigma_b})\tau^+_a \tau^+_b| 0_i^+ \rangle \\
{\cal M}_{T^{'}}&=&\langle 0_f^+| h_{T^{'}}\{(\vec{\sigma_a}\hat{\vec{r}}_{ab})
(\vec{\sigma_b}\hat{\vec{r}}_{ab})
-\frac{1}{3}(\vec{\sigma}_a
\vec{\sigma}_b)\} \tau^+_a \tau^+_b| 0_i^+ \rangle,\\
\alpha_1&=&\frac{T_1^{(3)}(0)G_V(1-2m_P(G_W/G_V))}{2 G_A^2 R m_e}.
\ea
$h_R$ and $h_{T^{'}}$ are neutrino potentials defined as
\ba{nepot01}
h_R&=&\frac{2}{\pi}\frac{R^2}{m_P}\int_0^\infty dq q^4
\frac{j_0(qr_{ab})f^2(q^2)}{\omega(\omega+\overline{E})}, \\
h_{T^{'}}&=&\frac{2}{\pi}\frac{R^2}{m_P}\int_0^\infty dq q^2
\frac{f^2(q^2)}{\omega(\omega+\overline{E})}
\{q^2 j_0(qr_{ab})
-3 \frac{q}{r_{ab}} j_1(qr_{ab})\}. 
\ea
Here $R$ denotes the nuclear radius, $m_P$ the proton mass.

Further $\omega=\sqrt{q^2+m_{\nu}^2}$, $q=|\vec{q}|$,
$\hat{r}=\vec{r}/r$ and $j_k(qr)$ are 
spherical Bessel functions. $(\omega+\overline{E})^{-1}$ is the energy 
denominator of the perturbation theory. The form factors 
$F^a_i(0)=F^a_i(q^2)/f(q^2)$ 
and $T_1^{(3)}(0)=T_1^{(3)}(q^2)/f(q^2)$
with $f(q^2)=(1+q^2/m_A^2)^{-2}$ 
($m_A^2=0.85$ GeV) 
have been calculated in the MIT bag model in \cite{adl}, $G_A \simeq 1.26$,
$G_V \simeq 1$ and
the strength of the induced weak magnetism 
$(G_W/G_V)=\frac{\mu_P-\mu_n}{2m_P}\simeq \frac{-3.7}{2m_P}$ is obtained by 
the 
CVC
hypothesis.

For comparison and to correct some details in \cite{Hir96} we first 
concentrate on the $S+P$ part,
, i.e. ${\cal M}_{T}=0$ in eq. \ref{t12}.
Inserting the numerical value of the matrix elements
${\cal M}_{GT^{'}}=2.95$ and ${\cal M}_{T^{'}}=0.224$ \cite{Hir96} 
for the special case of $^{76}$Ge 
and the half life limit obtained from the 
Heidelberg--Moscow experiment, $T_{1/2}^{0\nu\beta\beta}>1.2 \cdot 10^{25}y$, 
\cite{HM97,Pascos} one derives 
$\eta_q \leq 4.5 \cdot 10^{-8}$,
$\bar{\eta}_{(l)}\leq1.1 \cdot 10^{-8}$, 
corresponding to limits of
\ba{limits}
\lambda^{'}_{112}\lambda^{'}_{121}\leq 5.0 \cdot 10^{-6}
\Big( \frac{\Lambda_{SUSY}}{100 {\rm GeV}}\Big)^3 \nn \\
\lambda^{'}_{113}\lambda^{'}_{131}\leq 1.6 \cdot 10^{-7}
\Big( \frac{\Lambda_{SUSY}}{100 {\rm GeV}}\Big)^3 \nn \\
\Delta_n \lambda^{'}_{311}\lambda_{n13}\leq 9.9 \cdot 10^{-8}
\Big( \frac{\Lambda_{SUSY}}{100 {\rm GeV}}\Big)^3.
\ea
Here in the last equation
only the term corresponding to $\tilde{\tau}$ exchange has been kept.
These limits differ from the ones given in \cite{Hir96} due to 
an erroneous factor of $2$ in the definition of ${\cal M}_{S+P}$ 
in that reference as well as the improvement in the experimental bound.

Keeping the tensor part in eq. \ref{t12} the half life 
limit of the Heidelberg--Moscow Experiment implies:
\ba{limits2}
\lambda_{112}^{'} \lambda_{121}^{'}\leq  1.1 \cdot 10^{-6}
\Big( \frac{\Lambda_{SUSY}}{100 {\rm GeV}}\Big)^3, \nn \\
\lambda_{113}^{'} \lambda_{131}^{'}\leq 3.8 \cdot 10^{-8}
\Big( \frac{\Lambda_{SUSY}}{100 {\rm GeV}}\Big)^3 
\ea
for supersymmetric mass parameters of order 100 GeV.
For $\Lambda_{SUSY}\sim 1$ TeV, motivated by SUSY naturalness arguments, one 
obtains 
$\lambda_{112}^{'} \lambda_{121}^{'}\leq  1.1 \cdot 10^{-3}$,
$\lambda_{113}^{'} \lambda_{131}^{'}\leq 3.8 \cdot 10^{-5}$.
These are by more than a factor of four more stringent than the limits 
obtained from the scalar pseudoscalar part considered in \cite{Hir96}. 
The uncertainty of the nuclear matrix elements involved can be estimated to
be less than a factor of two. This is motivated by the fact that the
$2\nu\beta\beta$ halflife of $^{76}$Ge 
has been predicted correctly 
within a factor of 2 (the $2\nu\beta\beta$ matrix element within a factor of 
$\sqrt{2}$) \cite{2nu}. Since the uncertainty in both parts
($S+P$ and tensor) is expected to be about the
same, the improvement can still be considered as 
substantial. 

The obtained bound should be compared with the limits obtained from tree level
$K^0-\bar{K}^0$ and $B^0-\bar{B}^0$ mixing, which yield
$\lambda_{i12}^{'}\lambda_{i21}^{'}<1 \cdot 10^{-9}$
and
$\lambda_{i13}^{'}\lambda_{i31}^{'}<8 \cdot 10^{-8}$
respectively, for $m_{\tilde{e}}=100$ GeV \cite{bha}. While the
second generation is bounded by about three orders of magnitude more stringent
from the $K$ system, for the third generation the double beta bound is most 
stringent. Moreover, since the masses of different exchanged particles
(selectrons and squarks) enter, the limits are complementary in some sense.

In conclusion, we have performed a reanalysis of the SUSY--accompanied
neutrino exchange mode of neutrinoless double beta decay. Contrary to the
previous ansatz we included the tensor contribution to the decay rate,
which has been shown to be the dominant contribution.
This improves the limits on $\lambda_{11j}^{'} \lambda_{1j1}^{'}$ derived 
without the tensor contribution by a factor of four and thus provides the 
most stringent bound on $\lambda^{'}_{113}\lambda^{'}_{131}$.

\section*{Acknowledgement}
M.H. would like to acknowledge support by the European Union's 
TMR program under grant ERBFMBICT983000.

\newpage
\begin{figure}
\epsfysize=55mm
\epsfbox{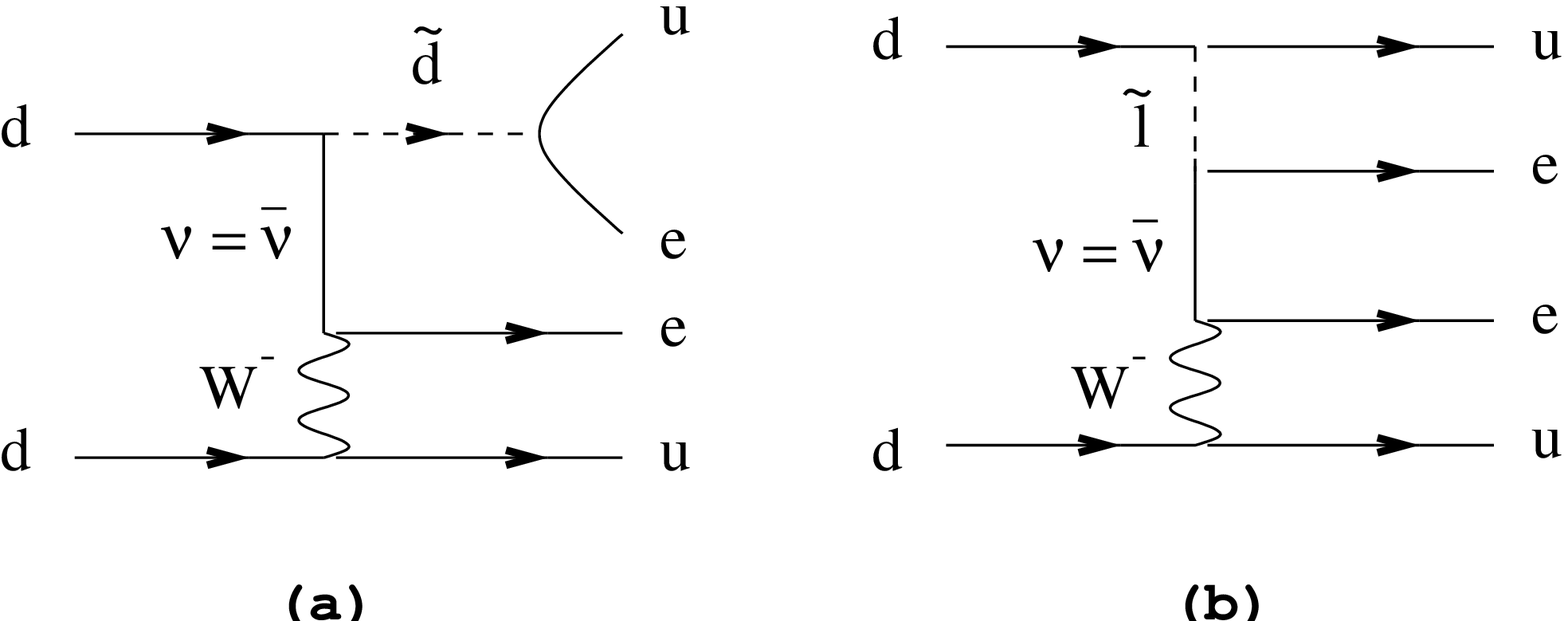}
\hspace{0.5 cm}
\caption{Feyman graphs for the neutrino exchange mechanism of neutrinoless 
double beta decay accompanied by (a) squark and (b) slepton exchange.}
\end{figure}


\begin{thebibliography}{99}

\bibitem{rpguts}
L. Hall, M. Suzuki, Nucl. Phys. B 231 (1984) 419; 
D. Brahm, L. Hall, Phys. Rev D 40 (1989) 2449;
K. Tamvakis Phys. Lett. B 382 (1996) 251;
G.F. Guidice, R. Rattazzi, hep-ph/9604339;
R. Barbieri, A. Strumia, Z. Berezhiani, hep-ph/9704275; K. Tamvakis, 
Phys. Lett B 383 (1996) 307; R. Hempfling, Nucl. Phys. B 478 (1996) 3;
A. Y. Smirnov, F. Vissani, Nucl. Phys. B 460 (1996) 37

\bibitem{rpss}
M.C. Bento, L. Hall, G.G. Ross, Nucl. Phys. B 292 (1987) 400;
N. Ganoulis, G. Lazarides, Q. Shafi, Nucl. Phys. B 323 (1989) 374;
A. Faraggi, Phys. Lett B 398 (1997) 95 

\bibitem{HERA}
C. Adloff {\it et al.} (H1 collab.), Z.Phys. C 74 (1997) 191;
J. Breitweg {\it et al.} (ZEUS collab.) Z.Phys. C 74 (1997) 207 


\bibitem{kalino}
J. Kalinowski, R. Rueckl, H. Spiesberger, P. M. Zerwas,
Z.Phys. C 74 (1997) 595

\bibitem{drei}
H. Dreiner,  in: "Perspectives on Supersymmetry", Edited
by G.L. Kane, World Scientific, Singapore; hep-ph/9707435

\bibitem{Kla97}
H.V. Klapdor--Kleingrothaus, in 
H.V. Klapdor--Kleingrothaus, H. P\"as (Eds.), Proc. Int. Conf.
{\it ``Beyond the Desert - Accelerator- and Non-Accelerator Approaches''},
Castle Ringberg, Germany, 1997, IOP, Bristol, 1998, p.485

\bibitem{Pascos}
 H.V. Klapdor-Kleingrothaus, H. P\"as,
 Proc. of the 6th Int. Symposium on Particles, Strings and Cosmology,
 (PASCOS98), Boston(MA), USA, March 22-27 1998, 
 to be publ. by World Scientific

\bibitem{mohorg}
R.N. Mohapatra, Phys. Rev. D 34 (1986) 3457;
J.D. Vergados, Phys. Lett. B 184 (1987) 55

\bibitem{Hir95}
M. Hirsch, H.V. Klapdor--Kleingrothaus, S.G. Kovalenko, Phys. Rev. Lett. 
75 (1995) 17, M. Hirsch, H.V. Klapdor--Kleingrothaus, S. Kovalenko, 
Phys. Rev. D 53 (1996) 1329


\bibitem{hir95d}
M. Hirsch, H.V. Klapdor--Kleingrothaus, S. Kovalenko, Phys. Lett. B 352
(1995) 1  


\bibitem{Hir96}
 M. Hirsch, H.V. Klapdor--Kleingrothaus, S.G. Kovalenko,
Phys. Lett. B 372 (1996) 181, Erratum: Phys. Lett. B 381 
(1996) 488 

\bibitem{Faes}
A. Faessler, S.G. Kovalenko, F. Simkovic, J. Schwieger
Phys. Rev. Lett. 78 (1997) 183

\bibitem{babu95}
K. S. Babu, R. N. Mohapatra, Phys. Rev. Lett. 75 (1995) 2276


\bibitem{Paes98a}
H. P\"as, M. Hirsch, S.G. Kovalenko, H.V. Klapdor--Kleingrothaus,
submitted to Phys. Lett. B


\bibitem{Doi85}
M. Doi, T. Kotani, E. Takasugi, Progr. Theor. Phys. Suppl. 83 (1985) 1

\bibitem{tom91}
T. Tomoda, Rep. Progr. Phys. 54 (1991) 53 

\bibitem{adl}
S. Adler {\it et al.}, Phys. Rev. D 11, (1975) 3309

\bibitem{HM97}
L. Baudis {\it et al.} (Heidelberg--Moscow collab.),
Phys. Lett. B 407 (1997) 219




\bibitem{2nu}
A. Balysh {\it et al.} (Heidelberg--Moscow collab.), Phys. Lett. B 322 (1994)
176;
K. Muto, E. Bender, H.V. Klapdor, Z. Phys. A 334 (1989) 177

\bibitem{bha}
G. Bhattacharyya, in \cite{Kla97}; we thank the author for bringing this bound 
to our attention

\end{thebibliography}
\end{document}